\documentclass[aps,prl,superscriptaddress,twocolumn,showpacs,amsmath,amssymb]{revtex4}

\usepackage{graphicx}
\usepackage{color}
\definecolor{DarkBlue}{rgb}{0.1,0.1,0.5}
\definecolor{Red}{rgb}{0.9,0.0,0.1}
\definecolor{Green}{rgb}{0.0,0.99,0.0}

\begin{document}

\title{Dependence of Band Renormalization Effect on the Number of Copper-oxide Layers in Tl-based Copper-oxide Superconductor using Angle-resolved Photoemission Spectroscopy}

\author{W. S. Lee}
\affiliation {SIMES, SLAC National Accelerator Laboratory, Menlo Park, CA 94025}

\author{K. Tanaka}
\affiliation {Advanced Light Source, Lawrence Berkeley Laboratory, Berkeley, CA }

\author{I. M. Vishik}
\affiliation {Department of Applied Physics, Stanford University, Stanford, CA 94305}

\author{D. H. Lu}
\affiliation {SSRL, SLAC National Accelerator Laboratory, Menlo Park, CA 94025}

\author{R. G. Moore}
\affiliation {SSRL, SLAC National Accelerator Laboratory, Menlo Park, CA 94025}

\author{H. Eisaki}
\affiliation {Nanoelectronic Research Institute, National Institute of Advanced Industrial Science and Technology, 1-1-1 central 2, Umezono, Tsukuba, Ibaraki, 305-8568, Japan}

\author{A. Iyo}
\affiliation {Nanoelectronic Research Institute, National Institute of Advanced Industrial Science and Technology, 1-1-1 central 2, Umezono, Tsukuba, Ibaraki, 305-8568, Japan}

\author{T. P. Devereaux}
\affiliation {SIMES, SLAC National Accelerator Laboratory, Menlo Park, CA 94025}

\author{Z. X. Shen}
\affiliation {SIMES, SLAC National Accelerator Laboratory, Menlo Park, CA 94025}
\affiliation {SSRL, SLAC National Accelerator Laboratory, Menlo Park, CA 94025}
\affiliation {Department of Applied Physics, Stanford University, Stanford, CA 94305}


\date{\today}

\begin{abstract}
Here we report the first angle-resolved photoemission measurement on nearly optimally-doped multi-layer Tl-based superconducting cuprates (Tl-2212 and Tl-1223) and a comparison study to single layer (Tl-2201) compound. A ``kink" in the band dispersion is found in all three compounds but exhibits different momentum dependence for the single layer and multi-layer compounds, reminiscent to that of Bi-based cuprates. This layer number dependent renormalization effect strongly implies that the spin resonance mode is unlikely responsible for the dramatic renormalization effect near the antinodal region.
\end{abstract}

\pacs{Valid PACS appear here}
\maketitle
The sharp renormalization effect observed by angle-resolved photoemission spectroscopy (ARPES) in the low energy excitation has drawn much attention \cite{Andrea03}. It is due to the coupling between electrons and some bosonic modes, which might be the long sought ``pairing glue" of Cooper pairs in high-$T_c$ cuprates, in analogous to the role of phonon in the conventional superconductors. Despite the progress made, it remains controversial whether the coupled bosonic modes are phonons or magnetic modes.  In particular, a signature of strong renormalization effect seen most clearly near the Brillouin zone boundary (referred as ``antinode" hereafter), in the form of broken-up dispersions and a ``peak-dip-hump" structure in the spectra, is taken as evidence for either coupling to neutron spin resonance  \cite{Kaminski01, Norman97, Sato03} or buckling phonon modes \cite{Cuk04,Devereaux04,Lee07_Bi2223}.  In principle, whether the renormalization effect exhibits a dependence on the number of CuO layer in a unit cell could shed light on this issue, as one expects the coupling to the buckling phonon to be dramatically enhanced in multi-layer compounds \cite{Devereaux04}.

ARPES studies on family of Bi-based cuprates (Bi$_2$Sr$_2$Ca$_{n-1}$Cu$_n$O$_{2n+4}$, n= 1, 2, and 3), by far the only testing ground for this issue, do not yield a conclusive result. The signature of a broken-up dispersion and peak-dip-hump structure in the single-layer compound is rather weak and appears abruptly in a narrow range of the momentum space with a different energy scale than that along the diagonal of the Brillouin zone (referred as ``node" hereafter)  \cite{Wei08}. This is in contrast to multi-layer compounds, in which a \emph{smooth} evolution from a dispersion kink to broken-up dispersion at a similar characteristic energy is seen \cite{Kaminski01,Cuk04,Sato03,Lee07_Bi2223}. This different momentum dependence between single and multi-layer compounds favors the interpretation of B$_{1g}$ buckling phonon; while the issue of a peak-dip-hump structure in the single-layer compound seems more uncertain, as the reported peak-dip-hump lineshape appears to be inconsistent among literatures \cite{Wei08, Meng08}. More importantly, neutron spin resonance measurement on the single layer compound, which is most decisive to distinguish these two scenarios, is not yet available.

Tl-based cuprates provide an opportunity to make significant progress on this issue as the spin resonance has been observed in the single layer Tl-based compound \cite{He02}.  The buckling phonon scenario and spin resonance scenario would lead to very different predictions: the spin resonance scenario shall predict a very similar result for all three compounds, whereas the buckling phonon scenario predicts distinct momentum dependence of the band renormalization between single- and multi-layer compounds. Unfortunately, due to the difficulties in growing high quality single crystals suitable for ARPES measurements, limited ARPES data are only available on heavily overdoped Tl-2201 \cite{Plate05, Peets07}; while the data on multi-layer Tl-cuprates have not yet been reported. In this letter, we report the first ARPES measurement on different members of the Tl-cuprate family near the optimally doped region, including the single layer (Tl-2201), bi-layer (Tl-2212), and tri-layer (Tl-1223) compounds. Our data not only confirm several general electronic properties of high-$T_c$ cuprates, but also strongly disfavor the spin resonance as the origin of antinodal renormalization effect.

Single crystals of nearly optimally-doped Tl$_2$Ba$_2$CaCu$_2$O$_8$ (Tl-2212), TlBa$_2$Ca$_2$Cu$_3$O$_9$ (Tl-1223) and slightly overdoped Tl$_2$Ba$_2$CuO$_6$ (Tl-2201) were grown using flux method. As-grown Tl-2212 ($T_c$ =107 K) and Tl-1223 ($T_c$ = 123K) crystals were chosen for the ARPES measurement. Tl-2201 crystals used in our measurement were prepared by annealing the as-grown crystal ($T_c$ $\sim$ 30 K) under a nitrogen flow at a temperature of 500$^\circ$C, yielding a T$_c$ of 80 K. The data were collected using a Scienta R4000 photoelectron spectrometer.  Measurements were performed at the SSRL beamline 5-4 using 28 eV photons and at the  ALS beamline 10.0.1 using 50 eV photons. The energy resolution was set at 15-20 meV for the data presented in this paper. Samples were cleaved and measured in the ultra-high vacuum ($ < 4\times10^{-11}$ Torr) to maintain a clean surface.

\begin{figure} [t]
\includegraphics [clip, height=3.0 in]{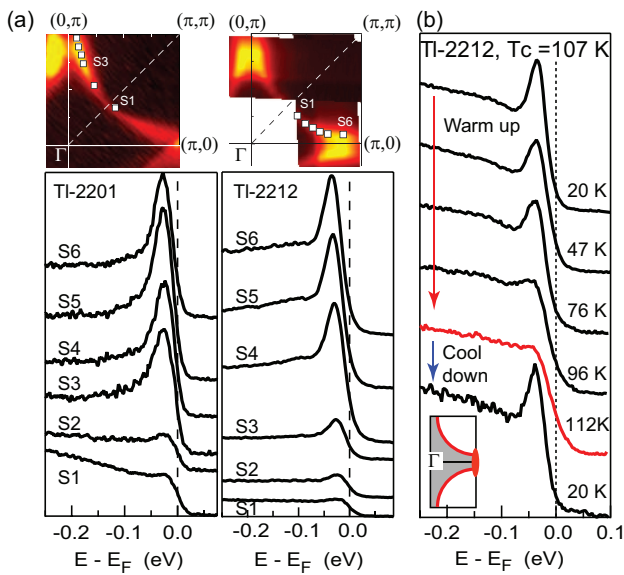}
\caption{\label{Fig1:FS_EDC} (a) The Fermi surface obtained by integrating spectra (taken at 20 K) within an energy window of $E_F \pm 10$ meV (upper panels) and their EDCs near $k_F$ along the Fermi surface (lower panels). The momentum positions of these EDCs were indicated by the square markers. All data were taken in the 2nd Brillouin zone and plotted in the 1st zone under the reduced zone scheme. The Fermi surface map of Tl-2212 is symmetrized with respect to the diagonal of the zone. (b) Integrated EDCs at the antinodal region of Tl-2212 (inset) in a cycle of temperature dependence measurements.}
\end{figure}

\begin{figure}[ht]
\includegraphics [clip, height=2.75 in ]{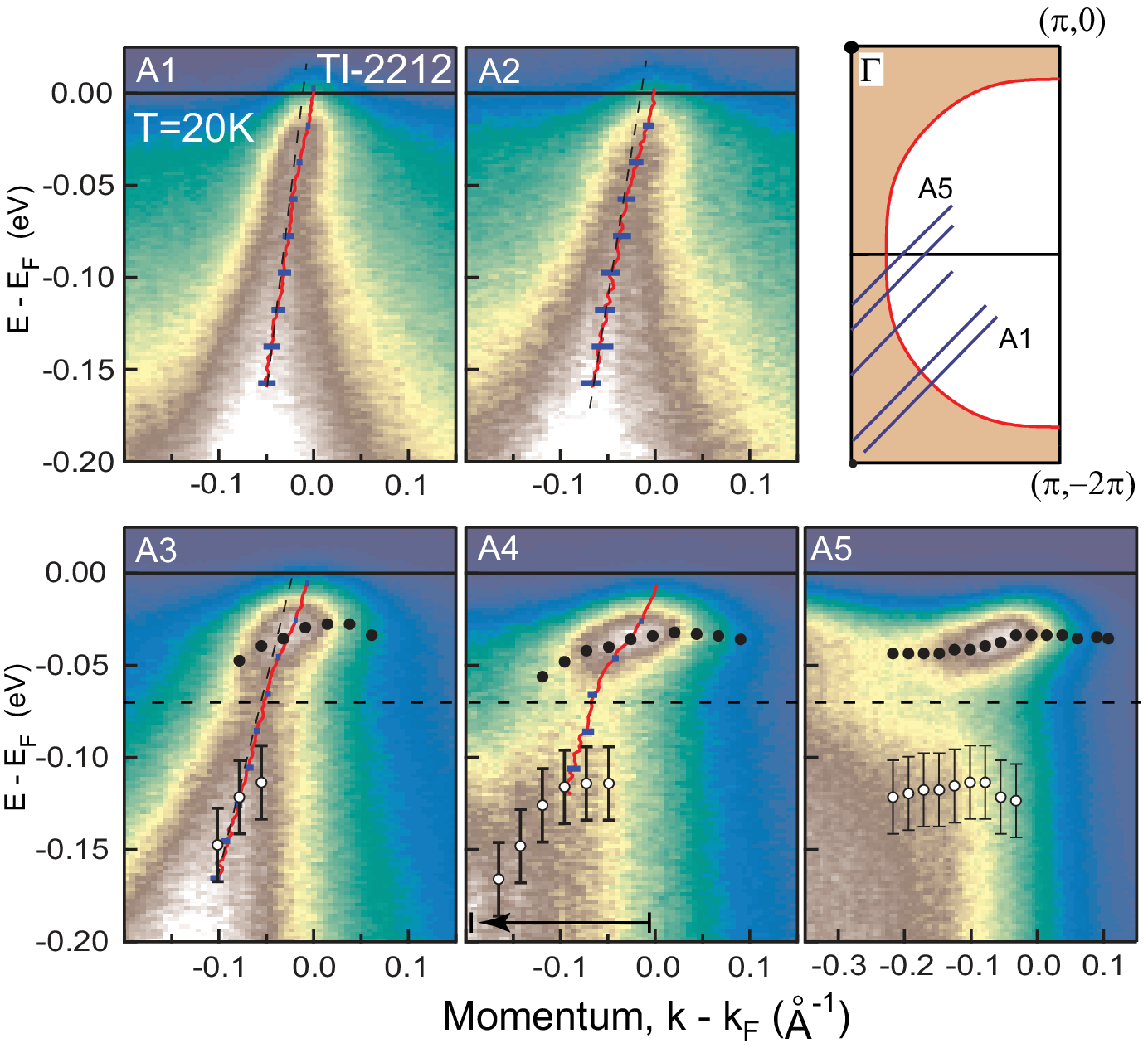}
\centering
\caption{\label{Fig2:Tl2212} Band dispersions of Tl-2212 near the Fermi surface along the cuts indicated in the upper-right panel. Data were taken at 20 K. Solid red curves are the band dispersion determined by the peak position of the momentum distribution curves (MDCs) via fitting to a Lorentzian function\cite{Kaminski01}. Blue short horizontal bars are the error bars at selected points represent the 3 $\sigma$ confidence levels from the fitted MDC peak positions. Black (white) markers are the band dispersion determined by the peak (hump) positions of representing EDCs. The bending back of the EDC-derived dispersion near $k_F$ is due to the opening of a superconducting gap. Black horizontal dashed line indicates an characteristic energy scale of the renormalization effect. Representing EDCs of A4 along the black horizontal arrow are plotted in Fig. \ref{Fig4:EDCs_Comp}(b)}
\end{figure}

Fermi surfaces of Tl-2201 and Tl-2212 are shown in Fig. \ref{Fig1:FS_EDC} along with the energy distribution curves (EDCs) at several different momentum positions. First, both Fermi surfaces consist of a hole-like sheet centered at ($\pi,\pi$). Sharp quasiparticle peaks exist near antinodal region for both compounds. This FS topology is consistent with that of other nearly optimally-doped cuprates \cite{Andrea03}. Second, the sharp quasiparticle peak becomes less pronounced when moving from the antinodal region toward the nodal region; near the nodal region, the ``peak" appears to be weaker and merges with a rising background at the higher binding energy. This behavior is opposite to the conventional expectation established by doping dependence studies on Bi-2212 and La$_{1-x}$Sr$_x$CuO$_4$ systems \cite{Andrea03, Yoshida03}, in which the quasiparticle peak is most well defined in the nodal region. We note that in heavily overdoped Tl-2201 system, a similar behavior of the quasiparticle peak has also been reported, and has been interpreted as an indication of a quantum critical point inside the superconducting dome, or as a result of strong forward scattering effect \cite{Plate05}. However, alternative explanations, such as the matrix element effect and the surface/crystal quality should not be excluded at this stage, and a further study is required. We also remark that although Fermi surface splitting has been observed in some multi-layer cuprates, such as Bi-2212 \cite{Feng01,Chuang01} and YBCO \cite{Borisenko06, Kaminski07}, we have not yet resolved a Fermi surface splitting in the optimally-doped Tl-2212.

\begin{figure}[t]
\includegraphics [clip, height=2.75 in]{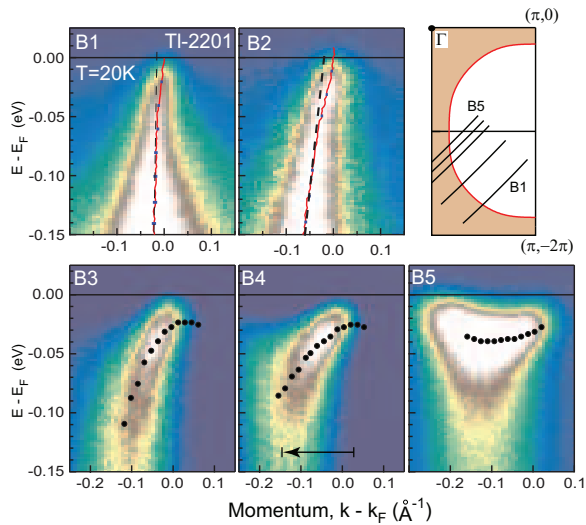}
\centering
\caption{\label{Fig3:Tl2201} Band dispersions of Tl-2201 near the Fermi surface along the cuts indicated in the upper-right panel. }
\end{figure}

In Fig. \ref{Fig1:FS_EDC} (b), integrated EDCs in the antinodal region in a cycle of temperature dependence measurements are displayed. The sharp peak gradually diminishes when the temperatures approaches $T_c$, and become undetectable at temperatures above $T_c$. The sample was then cooled back down to 20 K at which the sharp peak emerges again. This reproducibility of the sharp peak in the superconducting state proves that the emergence of a sharp antinodal peak below $T_c$ is indeed related to the onset of the superconductivity. We note that this observation is consistent with other cuprates with high $T_c$, such as optimally doped Bi-2212, Bi-2223 \cite{Andrea03}, and overdoped Tl-2201 with $T_c=74K$ \cite{Peets07}.  Our results confirm that the emergence of a sharp peak in the superconducting state is one of the universal properties of superconducting cuprates with high transition temperatures.

\begin{figure}[t]
\includegraphics [clip, width=3.25 in]{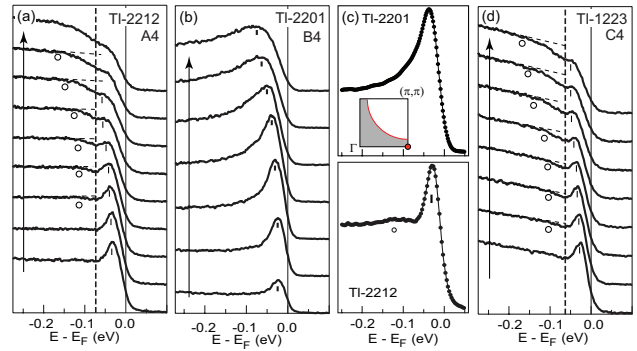}
\centering
\caption{\label{Fig4:EDCs_Comp} The representing EDCs of (a) A4 for Tl-2201, (b) B4 for Tl-2212, and (d) C4 for Tl-1223 along the black arrow indicated in Fig. \ref{Fig3:Tl2201}, \ref{Fig2:Tl2212}, and \ref{Fig5:Tl1223}, respectively. (c) EDCs for Tl-2201 and Tl-1223 near ($\pi$,0) (red point in the inset). Short vertical bars indicate the peak position. Open circles indicate the hump position. Vertical dashed line indicate the position of the ``dip". The dashed lines in (a) and (c) are guides-to-the-eye to make the ``hump" more discernible.}
\end{figure}

Next, we will discuss the momentum dependence of the band renormalization effect. In Fig. \ref{Fig2:Tl2212}, band dispersions of Tl-2212 at several different momentum positions on the Fermi surface are shown. A sudden change of the dispersion slopes, a kink at an energy of approximately 50-70 meV is observed near the nodal region (A1 and A2) .  When moving from the nodal region toward antinodal region (A3-A4), the ``kink" becomes more dramatic and eventually breaks the band dispersion causing an intensity depletion in the image plot of the spectrum at a characteristic energy of approximately 70 meV  (black dashed line). The representing EDCs of the broken-up band dispersion at A4 are plotted in Fig. \ref{Fig4:EDCs_Comp} (a).  When the band disperses toward higher binding energy (along the direction of the arrow), the sharp peak close to E$_F$ starts to diminish and the spectral lineshape at higher binding energy starts to deviate from a straight line, indicating convex curvature of representing a ``hump".

To emphasize the structure of the spectra, we simply define the peak position as the maximum of the sharp peak when it is still discernible in the spectrum, while hump position is set as the position where the spectrum starts to deviate from the apparent straight line drawn from the highest binding energy of our data. It is clear that the band dispersion breaks into two branches: a sharp peak branch (short bars) and a broad hump branch (open circles). Between these two features is a ``dip" in the EDC, which corresponds to the intensity depletion region seen in the image plot of spectrum (Fig. \ref{Fig2:Tl2212}). This broken-up dispersion and the peak-dip-hump structure are seen in a wide range of the Fermi surface: from approximately the midpoint between the node and antinode to the antinode as marked and displayed in A3-A5 of Fig. \ref{Fig2:Tl2212}. The fact that the dip is located at a similar energy scale (black dashed line in Fig. \ref{Fig2:Tl2212}) for A3-A5 suggests that the renormalization effect is dominated by one sharp bosonic mode in this momentum region.

Contrarily,  the single layer Tl-2201 system exhibits a qualitatively different momentum dependence of the renormalization effect. As shown in Fig. \ref{Fig3:Tl2201}, a ``kink" in the dispersion around 50-70 meV is also observed near the nodal region (B1 and B2). This dispersion kink becomes harder to be resolved near the antinodal region; as shown from B3 to B5  (from intermediate region to antinodal region), neither an apparent ''kink" in the dispersion nor an  intensity depletion in the image plot of the spectra can be discerned.  Indeed, as plotted in Fig. \ref{Fig4:EDCs_Comp} (b), the band dispersion at B4, which is located at a similar momentum position of A4, has only one branch consisting a peak structure with no sign of a peak-dip-hump structure as those observed in Tl-2212 compounds. The lack of a peak-dip-hump lineshape persists to the antinode as oppose to that of Tl-2212 (Fig. \ref{Fig4:EDCs_Comp} (c)).

\begin{figure}[t]
\includegraphics [clip, height=2.75 in]{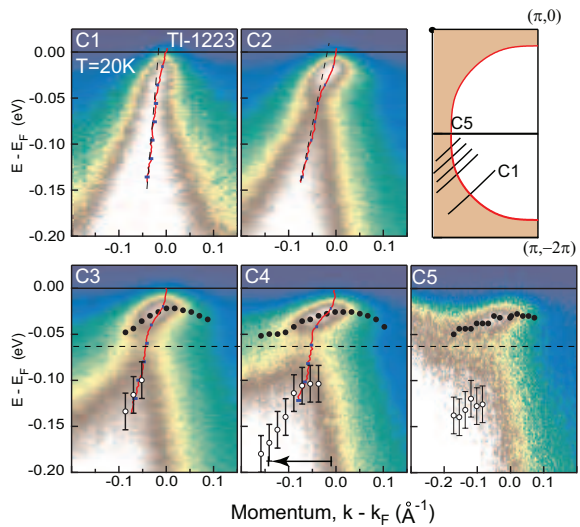}
\centering
\caption{\label{Fig5:Tl1223} Band dispersions of Tl-1223 near the Fermi surface along the cuts indicated in the upper-right panel. }
\end{figure}

To further verify the observed distinct momentum dependence of the renormalization effect between single layer and double layer Tl-based cuprates, we performed measurements on the Tl-based tri-layer cuprate, Tl-1223, which are shown in Fig. \ref{Fig5:Tl1223}. The momentum dependence of renormalization effect is found to be qualitatively similar to those observed in Tl-2212 (Fig. \ref{Fig2:Tl2212}); a broken-up dispersion with a dominant characteristic energy scale of 50-70 meV can be identified (black dashed line in Fig. \ref{Fig5:Tl1223}). As also shown in Fig. \ref{Fig4:EDCs_Comp} (d), the peak-dip-hump structure in EDCs is also seen in the similar momentum space range on fermi surface. We note that the smaller antinodal quasiparticle peak and higher spectral background observed in Tl-1223 system are perhaps due to the rough cleaved surface.  Tl-1223, unlike Tl-2201 and Tl-2212, does not have a natural cleaving plane in the crystal structure in which the Tl-O layer is shared by unit cells along c-axis.  We remark that the observed momentum dependent renormalization in multi-layer Tl-based compounds (Tl-2212 and Tl-1223) is similar to those observed in the optimally-doped multi-layer Bi-based cuprates, including Bi-2212 \cite{Andrea03,Cuk04} and Bi-2223 \cite{Lee07_Bi2223,Sato03}, where the renormalization effect along the entire Fermi surface is dominated by an energy scales near 70 meV in the superconducting state.

Therefore, the most intriguing finding of this work is the distinct momentum dependence of the renormalization effect between single layer and multi-layer Tl-based cuprates. This finding has an important implication on the origin of the mode responsible for the observed renormalization effect. The lack of a strong renormalization effect in the antinodal region in the single layer Tl-2201 system strongly suggests that spin resonance mode is \emph{not} the origin of band renormalization effect near the antinodal region, since the spin-resonance mode does exist in the Tl-2201 compound \cite{He02} and supposedly, should yield a strong renormalization near the antinodal region. This also casts strong doubts on the assignment of the antinodal region renormalization effect seen in Bi-2212 and Bi-2223 to the spin resonance mode \cite{Norman97,Kaminski01}.

C-axis phonons, on the contrary,  can behave very differently in single layer CuO$_2$ plane systems compared to multi-layer systems due to the environment surrounding the CuO$_2$ planes. For single layer materials, the CuO$_2$ plane lies in a mirror plane and thus c-axis vibrations of the CuO$_2$ plane can only couple to electrons to the second order of the atoms' displacements. Once the mirror plane symmetry is broken, such as in the multilayer systems, a coupling to the local c-axis field is possible at first order. In this way, the coupling of electrons to c-axis Raman modes (such as the A$_{1g}$ and B$_{1g}$ oxygen phonons) in single vs. multilayer cuprates are qualitatively different \cite{Devereaux95, Opel99}. In particular, it has been demonstrated that the B$_{1g}$ buckling phonon is capable of producing the renormalization feature observed in multi-layer compounds \cite{Devereaux04}, which best explains the layer number dependent renormalization observed in family of Tl-cuprates. Finally, we remark that this B$_{1g}$ buckling model favors to form Cooper pairs in the $d$-wave channel \cite{Devereaux95}, which could be one of the factors to induce a higher $T_c$ in multi-layer systems.

SSRL is operated by the DOE Office of Basic Energy Science, Division of Chemical Science and Material Science. This work is supported by DOE Office of Science, Division of Materials Science, with contract DE-AC02-76SF00515

\end{document}